# CHALLENGES AND OPPORTUNITIES TO ASSURE FUTURE MANUFACTURING OF MAGNET CONDUCTORS FOR THE ACCELERATOR SECTOR

White Paper for the Accelerator Frontier
Snowmass '21

Lance Cooley[1,2], David Larbalestier[1,2], and Kathleen Amm[3]

1. Applied Superconductivity Center (ASC), National High Magnetic Field Laboratory (NHMFL), Florida State University
2. College of Engineering, Department of Mechanical Engineering, Florida A&M University and Florida State University
3. Superconducting Magnet Division, Brookhaven National Laboratory (BNL)

## Executive Summary

*$Nb_3Sn$ magnet conductors will continue to be the workhorse material for accelerator magnets over the coming decade because they can deliver significantly higher magnetic fields than Nb-Ti at significantly lower cost than higher performance HTS conductors. High current-density $Nb_3Sn$ conductor suitable for present accelerator magnets for the High-Luminosity LHC (Hi-Lumi) upgrade, i.e. ~12 T field, is commercially produced in long lengths in a mature fabrication process. R&D enhancements for "advanced $Nb_3Sn$" could improve performance by 30% or more in the 15-16 T field range envisioned for future dipoles, and conductors could become available in 2–5 years. New developments in cabled REBCO tape HTS conductors and 2212 round-wire strand are creating opportunities for hybrid $Nb_3Sn$-HTS dipole magnets approaching 20 T and solenoid magnets pushing toward 50 T. Disruption of Nb-Ti and $Nb_3Sn$ magnet technology and increasing cross-over points for hybrid and stand-alone HTS options operating above 4 K temperature should be expected as HTS magnet technology continues to develop and mature.*

*No superconductor is presently manufactured at the huge scale required for a high energy or muon collider. MRI Nb-Ti is the only present tonnage conductor, but its architecture is simpler than any envisaged accelerator conductor. Moreover, it is made at very small margin because of continuing large cost pressures from the MRI magnet makers. The superconductor industry of today lacks adequate economic margins from Nb-Ti MRI wire manufacture to support development of innovations such as advanced $Nb_3Sn$ and HTS conductors. Market pull for the Hi-Lumi grade of $Nb_3Sn$ exists is presently insufficient to drive manufacturing scale beyond the 30-ton Hi-Lumi procurement.*

*Accelerator magnets require premium grades of $Nb_3Sn$ conductor that have a limited supplier base. For example, the $Nb_3Sn$ conductor used for the High-Luminosity LHC has only one supplier that can meet the required performance and piece length, whereas the larger procurement activity for ITER was sourced from 8 suppliers. With Hi-Lumi coming to an end, there is significant concern about how to keep premium $Nb_3Sn$ manufacturing active enough to prevent loss of expertise and atrophy of capabilities. It is our view that present public-private partnerships must be enhanced. Annual procurements of significant scale and conductor stockpiling will be necessary to maintain pre-production readiness of existing conductor designs. Re-thinking the innovation network will be needed to facilitate conductor innovation in view of the tension between large-volume, low-margin tonnage Nb-Ti and the smaller-scale, Hi-Lumi $Nb_3Sn$ with limited end-use pull, and developmental $Nb_3Sn$ and HTS conductors which cannot be funded by present industry profit margins.*

*Broader coordination between federal offices and the conductor-magnet industry ecosystem could encourage the emergence of marketplace demand that results in both better and lower cost conductors. Synergies are envisioned between a sustained development program for a muon collider and technology maturation for emerging industries since both entities require high-field solenoids. Increased partnership between accelerator sector magnet developers and magnet technology for industry could provide important bridges across development obstacles by making national laboratory infrastructure available to companies. The reshaping of national agendas to address climate and energy sustainability is instigating new applications in fusion and electric machines with cross-fertilization to accelerators. Transformations in health and medicine also have potential overlap with conductor and magnet technology in the accelerator sector. A portfolio of industries could be nurtured by the accelerator sector via innovation institutes and magnet technology centers of excellence to serve national and economic security needs and advance future accelerator technology needs.*

## Development path of magnet conductors from small samples to large-scale production

### Basic requirements at small-sample scale

Superconducting materials that could eventually become important conductors start out from discoveries with small samples, e.g., pellets, thin films, small cast or melted rods, etc. Among the starting points of today's conductors are arc-melted alloys of Nb, sintered powders of $MgB_2$ and $(Ba_xK_{1-x})_2FeAs$, and films of $YBa_2Cu_3O_7$ (YBCO, or more generally REBCO where RE= Y or other rare earth element). Measurements of certain basic properties reveal whether the material has potential to impact magnet technology. Among the first tests of a new material are the determination of basic attributes including:
1. Whether there is a high enough upper critical field or irreversibility field at accelerator operating temperature (usually 4 K). See Figure 1.
2. Whether high macroscopic (across many grains) and microscopic (within one grain or crystal) critical current density $J_c$ is present. See Figure 2.

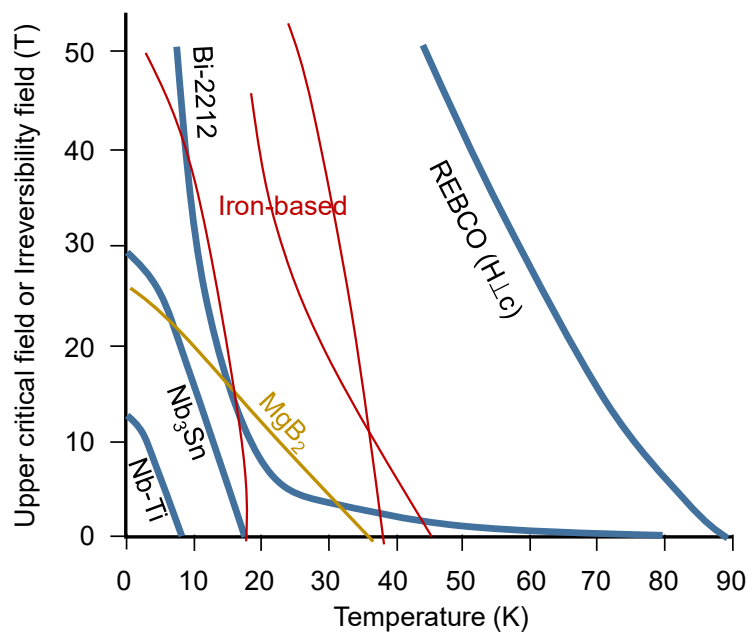

**Figure 1:** A plot of upper critical field or irreversibility field vs. temperature for magnet conductors used by the accelerator sector and potential new conductors.

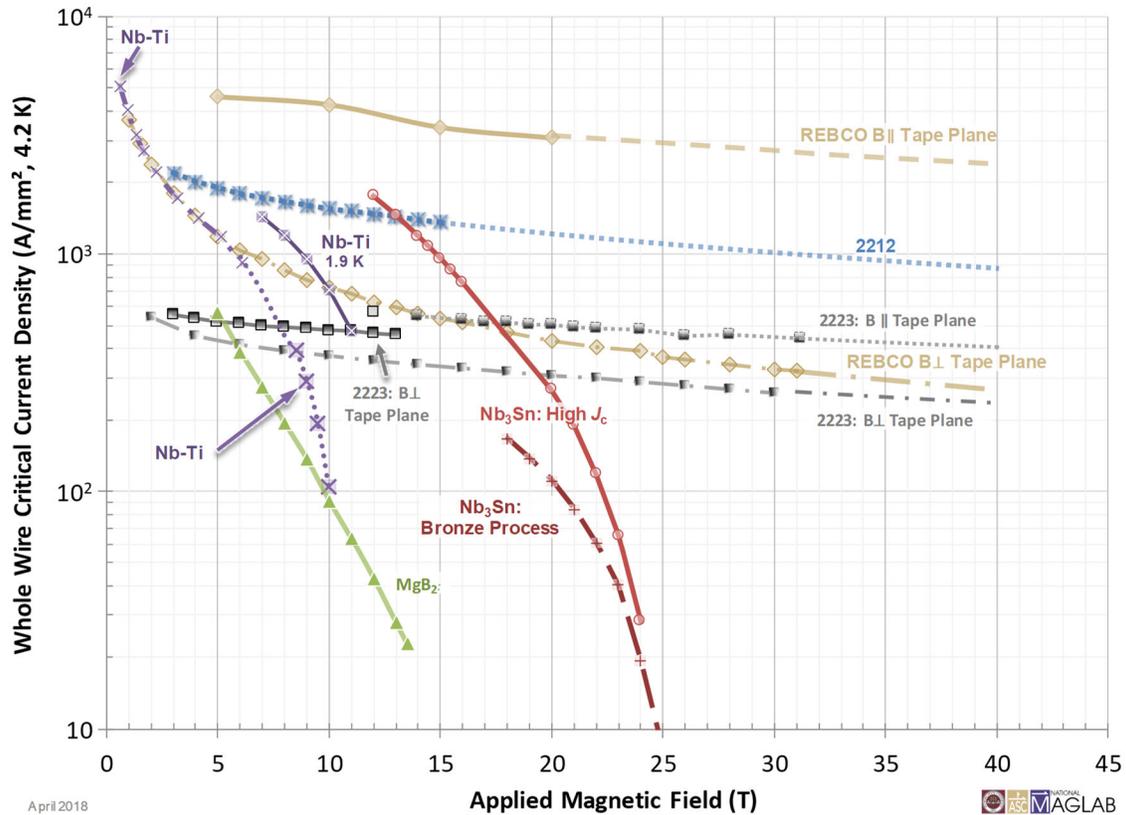

**Figure 2:** Critical current density of industrial wires and tapes used in superconducting magnets. Plot is maintained by Peter Lee, see https://nationalmaglab.org/magnet-development/applied-superconductivity-center/plots.

3. When blockage of current is evident, <u>understanding the cause of the connectivity obstruction</u> must be attained, so that methods of enhancing that connectivity can be developed (for example, by melt texturing of Bi-2212 ($Bi_2Sr_2CaCu_2O_8$) or the biaxial growth of REBCO on textured templates).
4. <u>Whether the synthesis methods are compatible with parallel-bonded high conductivity metals,</u> especially copper.
5. <u>Whether there is significant crystalline and property anisotropy</u>, which may strongly impact vortex pinning and mechanical properties.
6. <u>Whether there are ready means to increase flux pinning by suitable additives and mitigate connectivity limitations</u> by for example grain boundary engineering.

Statement: Support of materials science research that allows the HEP community to capitalize on new materials and enables their application to provide technological breakthroughs is and has been vital to the development of superconducting applications.

### Considerations for research on nascent wires

Research to develop a new material in a wire or tape form takes place using billets or stacks that are up to about 2 kg mass and lengths ranging from a few meters to 100 m. These ranges are tied to the capabilities for development in university, SBIR, and industry research settings. To perform the first stages of wire development, requirements *in addition to those for basic material studies above* are:

1. Compatibility with copper wire fabrication methods, i.e. wire or tube drawing to large strain. *Composite* wire manufacturing approaches include combinations of ductile metals (e.g. Cu + Nb-Ti alloy, Cu + Nb + Sn for later reaction to form brittle $Nb_3Sn$) and combinations of packed powders in ductile tubes (e.g. Cu or Cu-alloy tube + $MgB_2$ powder, Ag tube + Bi-2212 powder, possibly Cu tube + powder of Fe-based superconductor).
2. If not, then compatibility with long-length fabrication by other means, e.g. reel-to-reel deposition on (REBCO and its buffer and template layers on Hastelloy or stainless steel) or printing. The 2D conductors emerging from such processes complicates magnet technology.
3. Understanding how processing serves to define properties is the basic materials science paradigm that drives all optimization. The materials science involved in the discovery of a promising superconductor can be very different than the materials science that goes into production of viable magnet conductors capable of being fabricated in long lengths. All processing steps must serve the optimization of structure and thereby optimization of properties—this was the key lesson from the development of Nb-Ti magnet conductors.
4. Development of heat treatments and other reaction processes that are compatible with magnets. Essentially all materials with operational envelopes beyond Nb-Ti are brittle, meaning that they must be made from ductile components that are placed into a furnace to allow the superconductor to form *in-situ* via reaction, e.g. $3Nb + Sn \rightarrow Nb_3Sn$. To avoid cracking of the brittle material, magnets are usually fabricated by "wind and react" methods, which implies that the conditions for the superconductor formation reaction must also be compatible with the materials used in the magnet. This includes compatibility with insulation and with mechanical and structural components. $Nb_3Sn$ and Bi-2212 are generally used in a wind-and-react approach.
5. Understanding of how the materials depend on strain, both reversibly and irreversibly. These properties must be understood at the basic superconducting materials level and at the conductor level where normal metal conductor such as Cu or a strong substrate like Hastelloy may be present. The strain limits must be determined at the whole-conductor level, and these may enable alternative "react and wind" magnet strategies, where the conductor is reacted to form the brittle superconductor on a spool and subsequent magnet winding avoids strain above the threshold for cracking (REBCO is always supplied in this final superconducting state).
6. If the superconductor is not isotropic and is formed with texture on a tape substrate (e.g. REBCO), then understanding of the variation of properties with orientation is essential.

## Considerations for industry R&D and scale-up

Transfer of nascent wire ideas to industry comes with expectations of a product and viability of a marketplace. Trials scale to the size of partial and full billets, for instance 10, 20, and 45 kg for $Nb_3Sn$ corresponding approximately to 1–10 km single wire piece lengths. Basic attributes of a product need to emerge from industrial R&D, including:

1. Compatibility with existing manufacturing infrastructure to avoid unnecessary new investment in infrastructure that would generate a wider "valley of death" before adoption.
2. Understanding requirements for specialty manufacturing. Some manufacturing paths require techniques that do not overlap the art of copper wire manufacturing. Examples include the special $NbSn_2$ powder manufacture developed for powder-in-tube (PIT) $Nb_3Sn$, vapor deposition methods, specialty shape forming, and additive methods. Specialty manufacturing requirements can be a significant cost detriment, as for example for PIT $Nb_3Sn$.
3. Flexibility of conductor architecture and opportunity for innovation. Innovation cycles of "better, faster, cheaper" greatly help both users and manufacturers.
4. Scaling to long length with high yield, low breakage, and low rejection rates. Product margin must cover capital, operating, and materials expenses associated with unsellable pieces not suitable for customer use.

5. <u>Consistency of superconducting and mechanical properties</u> along the length of single pieces, from piece to piece, and from production batch to batch.
6. <u>Consistency of key performance parameters</u> such as critical current along the length of single pieces, from piece to piece, and from batch to batch. For $Nb_3Sn$, conductor optimization has at least two performance parameters: critical current and copper conductivity.
7. <u>Development of measurement and testing capabilities needed to unambiguously validate key properties.</u>

## Considerations for industry production readiness

The full production run for a future accelerator facility could require procurement of hundreds or thousands of tons of magnet conductor over a period of several years. Production readiness requires:
1. <u>Determination of production specifications and related quality assurance.</u>
2. <u>Demonstrated ability to deliver to specifications with low reject rate</u>.
3. <u>Successful delivery of orders of significant size sustained over long duration</u>. Annual procurements of 100–1000 kg were sustained over a decade during LARP [1], establishing manufacturing capability and generating much-needed production and delivery statistics. Similar procurements preceded the major production runs for the LHC, ITER, SSC, and RHIC.
4. <u>Sufficient information to arrive at a procurement cost estimate</u>.
5. <u>Availability of test facilities and inter-laboratory benchmarks or standards to verify the above</u>.

> Statement: Driven by magnet development needs at national laboratories, collaborations between fundamental materials researchers in universities and labs with the conductor manufacturers has produced a 40-year long virtuous cycle of continuous conductor development for the accelerator sector. Sustained support for this "secret sauce" is vital to optimization of future conductors too.

> Statement: Sustained large procurements of conductor are vital to keep manufacturing capability warm, encourage retention of know-how and workforce, validate expectations for a production run, and encourage growth in the related non-research marketplace.

> Statement: Discussion about conductor cost should not begin until production readiness has been achieved and supply chains are established.

> Statement: Support of standards and test facilities is crucial for industries to quantify performance and assess technology and manufacturing advancement.

## Magnet conductors in the Accelerator Sector portfolio for the next decade

Table I summarizes in approximate terms the conductors available for procurement in 1 km lengths that meet the requirements of accelerator magnets. For example, $Nb_3Sn$ conductor designs for the High-Luminosity LHC upgrade ("Hi-Lumi LHC") are referenced but those for fusion are omitted, because fusion designs are optimized with parameters unsuitable for high-field accelerator magnets. The conductor development life cycle described above occurs over ~1 decade of time, so the position of $Nb_3Sn$ as the workhorse conductor and HTS as contributing conductors should only be expected to change slowly. We comment about potential disruptive factors, which could change the table substantially, at the end of this document. Several $Nb_3Sn$ conductors under development, e.g. PIT and Hf-alloyed variants, have the potential to reach 1 km length and would be expected to have manufacturing profiles like the production conductors shown below. A comprehensive review of development activity led by CERN was published recently [2]. $MgB_2$ and Bi-2223 conductors are included because of their potential role in supporting

| Table I: Magnet conductors available for procurement in length > 1 km ||||||
|---|---|---|---|---|
| SC material | Billet or batch mass | Annual production | Relative cost | Comments |
| Nb-Ti | 200-400 kg | Hundreds of tons | 1 | Driven by MRI industry |
| $Nb_3Sn$ RRP | 45 kg | 5–10 tons | 5 | Driven by general purpose and NMR magnets and by Hi-Lumi LHC |
| $Nb_3Sn$ PIT | 45 kg | < 1 ton | 8 | Cheaper RRP is also generally more capable |
| Bi-2212 | 20 kg | < 1 ton | 20–50 | See note (1). |
| REBCO | 10 kg | < 1 ton; few tons for fusion | 20–50 | See note (2). |
| Bi-2223 | 20 kg | < 1 ton | 20–30 | Current leads (3). |
| $MgB_2$ | 20 kg | < 1 ton | 2 [5] | Current transfer cables feeding magnets. |

Notes:
(1) Costs for Bi-2212 are artificially high because full-time R&D teams are supported by only a few conductor orders per year.
(2) Privately funded fusion projects, which target e.g. 20 T at 20 K, envision magnets that do not require large stores of helium. This could make REBCO the choice for the accelerator sector if avoidance of helium becomes a priority.
(3) Although Bi-2223 (i.e. conductor based on $Bi_2Sr_2Ca_2Cu_3O_{14}$) has current density that is at present significantly lower than that of either Bi-2212 and REBCO, Bi-2223 is unique for bridging temperature zones from 110 K downward, such as for magnet current leads.

accelerator magnets, e.g. like has been done for the CERN Superconducting Link [3], and in niche magnets where heat load is important [4].

> Statement: End-use pull is the most significant factor that drives down cost and motivates production scale-up. End-use pull from marketplace products is weak for $Nb_3Sn$ compared to Nb-Ti. Commercial HTS magnet technology is just beginning and synergistic development with the accelerator sector is now starting, especially through SBIR and STTR vehicles.

### Restacked-rod-process RRP $Nb_3Sn$ conductor: the present workhorse conductor

The conductor produced for the Hi-Lumi LHC, shown in Figure 3, uses an architecture called "restacked rod process" [7],[8] wherein monofilament rods of Nb or Nb alloy are sheathed in copper, extruded, drawn, and stacked in an annulus around a central Cu core, wrapped with a diffusion barrier, and extruded a second time. Part of the core is then replaced with tin, and this assembly forms what is called a sub-element. After further drawing, the sub-elements are stacked (i.e. the original rods are re-stacked) and further drawing proceeds to the final size. A reaction sequence produces mixed Cu-Sn phases at low to moderate temperatures (below 400 °C) before a final high-temperature segment facilitates a diffusion reaction between Cu-Sn and Nb (or Nb-alloy) to form Nb$_3$Sn (or Nb$_3$Sn alloyed with Ti, Ta, etc.) [9].

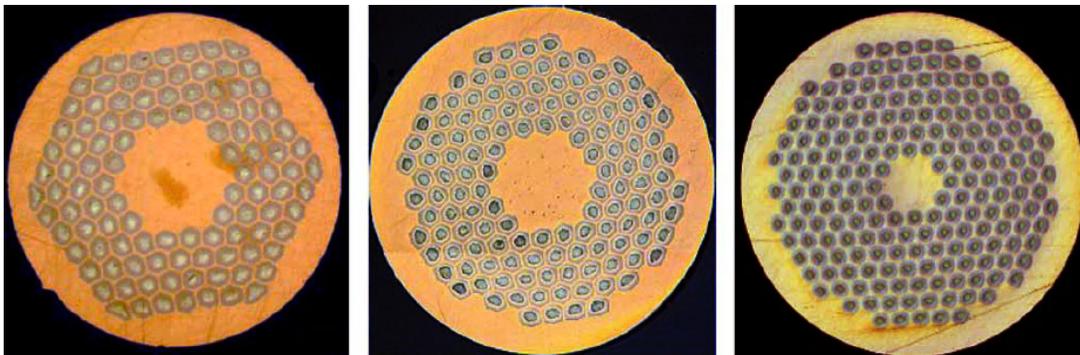

**Figure 3**: Cross-section of different 0.85 mm diameter Hi-Lumi LHC quadrupole conductors: At left is a 108/127 RRP strand from Bruker-OST, center is a 132/169 RRP strand from Bruker-OST, and at right is a 192/217 PIT strand from Bruker-EAS. Images were supplied by the manufacturers to the upgrade project as part of quality control. See Ferracin *et al.* [6].

Conductor naming typically refers to the number of sub-elements that occupy a theoretical number of sub-element sites while keeping hexagonal symmetry. The theoretical number obeys the formula $3n(n–1) + 1$ for $n$ layers around a central rod (the first layer), i.e. 1, 7, 19, 37, 61, 91, 127, 169, 217, 271, … for $n$ = 1 to 10. The number $n$ also identifies the number of hexagonal units along the flat outer edge of the stack. The Hi-Lumi quadrupole strand used 108 of 127 possible sites, and so was identified as "108/127", with the central 19 sites being copper. Research strands leading up to the Hi-Lumi LHC project explored configurations restacks from 61 up to 217 sub-elements, including copper [10],[11]. Procurements in 2021-2022 specify 150/169 and 162/169 designs and 1.0 to 1.1 mm final diameter.

A key limitation on reducing hysteretic losses and minimizing field errors due to large magnetization currents is presented by the fact that the final reaction stage grows the individual Nb filaments into a solid Nb$_3$Sn annulus that makes the effective filament diameter $d_{eff}$ that of the whole connected annulus rather than the starting Nb filament unless countermeasures are taken to divide the annulus, e.g. by substituting pure Ta rods [12]. Here, the product $J_c d_{eff}$ is associated with energy contained in the strand magnetization, which can be uncontrollably released as a flux jump and potentially initiate a magnet quench. The effective filament diameter is approximately given by

$$d_{\text{eff}} = d_w [N(1+R)]^{-1/2}$$

where $d_w$ is the wire diameter, $N$ is the number of sub-elements and $R$ is the ratio of stabilizer copper area to total sub-element area. Flux jumps are serious challenges for $d_{\text{eff}} > 60$ μm. For the 0.85 mm diameter Hi-Lumi quadrupole strand, $d_{\text{eff}}$ is nominally 55 μm. Re-stacking a larger number of sub-elements to make $d_{eff}$ smaller, increasing $N$, however, works in opposition to manufacturing yield, uniformity, performance, and cost. These trade-offs presently dictate present RRP design optimizations. Cables for future dipoles using present RRP designs will need larger diameter strands to carry more current, with present procurements seeking 1.0 to 1.2 mm diameter. Keeping $d_{\text{eff}}$ at ~60 μm requires re-stacks with $N$~200. Higher $J_c$ Nb$_3$Sn conductors would allow smaller wires, adding flexibility into the RRP strand design and also the fabrication of cables made from them.

The RRP design has proven to be quite flexible in over a decade of development. The tin insertion step and the stacking pattern in the annulus provides the means to vary the ratio of Nb to Sn, thereby tuning the Nb$_3$Sn reaction dynamics. Higher Sn content, e.g. a Nb:Sn ratio of 3.4:1, creates higher tin activity and drives the average Nb$_3$Sn composition across the layer closer to stoichiometry for a typical final reaction stage, e.g. 665 °C for 48 hours. This results in higher irreversibility field and $J_c$. However, aggressive reactions can drive Sn through the diffusion barrier and into the copper. Tin is a strong electron scatterer in copper, and even as little as 0.1% Sn can reduce the conductivity, as measured by the residual resistance ratio RRR, by a factor of 10 [13]. The resulting contamination adversely affects quench protection. A reduction in the availability of Sn, by using Nb:Sn ratios of 3.5:1 to 3.6:1, results in better ability to aggressively drive the diffusion reaction with less risk of reducing RRR [11]. This gives better multi-parameter optimization [1]. Figure 4 shows the outcome of this change for the Hi-Lumi LHC production run, which used a ratio of 3.6:1.

Adjustments of the architecture provide resilience against damage due to cabling. The process to make Rutherford cables requires rolling through a Turk's head where sharp bends at the cable edges lock in the strand geometry. Cabling research, which is not discussed at length in this document, typically probes the limit of compaction where strands are locked into position, but the performance is not seriously degraded. The RRP architecture facilitates some degree of adjustment to accommodate deformation of sub-elements during cabling. However, excessive local shear deformation of sub-elements can break diffusion barriers and allow Sn to have unobstructed access to the copper, usually resulting in significant loss of RRR. For Hi-Lumi quadrupole strands, rolling of round test strands to 85% of the initial diameter produced properties that tracked well with the degraded properties observed from strands extracted from cables, where in particular the RRR of the rolled strand was less than that measured around the sharp bends in cables [14],[15]. Rolled strand experiments thus provided a conservative predictor of cable performance. Also important is that the critical current of RRP strands is quite robust against rolling or cabling, because for reasons not fully understood the Nb-Sn diffusion reaction converts nearly as much of the sub-element area to $Nb_3Sn$ in distorted sub-elements as for sub-elements free from distortion.

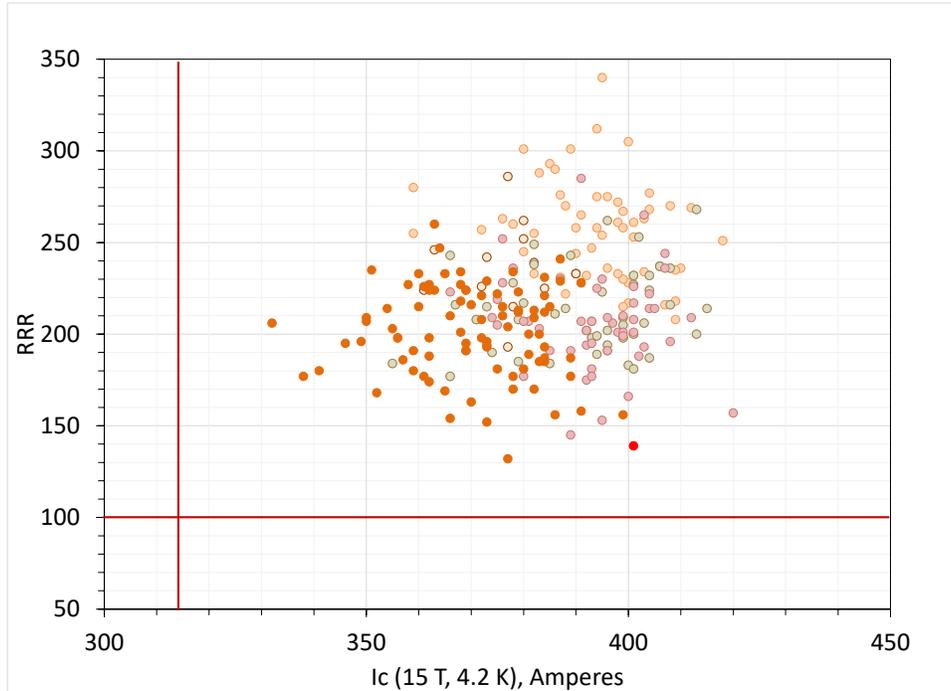

**Figure 4:** Cumulative data for RRP strand delivered for the US contribution to the Hi-Lumi LHC upgrade is represented in this plot of the RRR vs wire critical current at 15 T, 4.2 K. Each dot represents a quality control sample tested by the manufacturer, and the different shading identifies different procurement stages over the entire run. Importantly, these data are for strand rolled to reduce the diameter by 15%, which was found to simulate very well the degradation at sharp bends in Rutherford cables. Vertical and horizontal lines define the specifications for rolled strand.

The continued improvement of RRP conductors will be vital to future colliders. Two sections below address activities that crosscut all $Nb_3Sn$ conductor designs: (a) replacement of Pure Nb filaments by Nb-alloys; and (b) incorporating high heat-capacity materials into the strand architecture. In addition to these opportunities, improvement of RRP conductors also needs research attention given to:

- Understanding better how the sequence of Cu-Sn and Cu-Sn-Nb phases that evolve during the reaction heat treatment might be optimized for better $Nb_3Sn$ performance at >15 T. Prior work [16],[17] identified opportunities that that have not yet been fully exploited.
- Reducing $d_{eff}$ without concomitant loss of $J_c$ performance, loss of yield, or increase of cost.
- Scaling production to 100 kg billets or larger.
- Modifying the conductor architecture and heat treatment to optimize properties above 15 T.

### Powder-in-tube PIT conductor and rod-in-tube RIT conductors

Tube-based $Nb_3Sn$ approaches replace the annulus of stacked rods in RRP conductors with a solid tube of Nb or Nb-alloy. In principle, this removes one stage from the fabrication process, i.e. RRP: monofilament Cu/Nb, sub-element annulus, finished conductor; PIT: packed mono-tubes, finished conductor. The tin source, often combined with copper, can be a powder or a rod inserted in the center of the tube, whereby a radial diffusion reaction is used to form the $Nb_3Sn$ layer. The assembled tube and core material is restacked in a similar fashion as described for RRP conductors, where design flexibility is provided by tube shape (hexagonal or round), separation, and composition of components in the tube core. Reports [18] investigated restacks with as many as 744 tube sub-elements ($n = 18$ with 169 sub-elements being Cu) at tube diameter approaching 15 µm for 0.7 mm wire diameter.

Tubes are sourced as a raw material, which means that conductor manufacturing must extend from the art of composite wire drawing to also encompass the art of seamless tube manufacture. Piercing plates followed by deep drawing or spinning, dynamic flow forming, hydroforming, electroforming, and seamless mandrel extrusion are established methods for tube manufacture. Tubular conductors can in principle be made both with powder inside the tube [19],[20] as in the PIT conductors made by Bruker-EAS using $NbSn_2$ powder [19], or with solid rods, as for the conductors rod-in-tube conductors made by HyperTech Research [20]. More than a decade of development both for NMR magnet and for Hi-Lumi have shown it to be a flexible architecture but also one which is ultimately inferior to RRP architectures. An important mechanical concern is that all powder conductors require powder sliding during wire fabrication and this cannot be accomplished with full density cores. Typical residual porosities are about 1/3 which leads to greater irreversible strain sensitivity than in full density RRP wires [21],[22],[23]. A second factor is that the reaction path from the $NbSn_2$ source to $Nb_3Sn$ goes through "Nausite" $Nb_{0.25}Cu_{0.75}Sn_2$ and $Nb_6Sn_5$ which forms as a very large grain phase that rejects Cu into its grain boundaries when finally converting to very large grain $Nb_3Sn$ with poor vortex pinning, impaired connectivity and low $J_c$ properties [24],[25],[26],[27]. Thus the highest $J_c$ values of PIT have always been inferior to those of RRP conductor when expressed in the important conductor metric of $J_c$ equal to the critical current divided by the area of all phases including the diffusion barrier needed to make the $Nb_3Sn$.

In addition to these important areas, and in addition to the research topics discussed for RRP conductors, further improvement of tube-type $Nb_3Sn$ conductors could result from:
- Improved manufacturing experience with tubes
- Improved manufacturing of fine powders of Nb-Sn and Cu-Sn intermetallic compounds
- Development of full density routes using ductile Cu and Sn components
- Continued use in magnets to reveal conductor vulnerabilities

### Other Nb₃Sn designs

A significant advantage of the RRP and PIT conductor designs above is the placement of tin, generally as a Cu-Sn phase, next to the niobium in sub-elements. This creates high tin activity, which leads to rapid reaction and smaller composition gradients in the $Nb_3Sn$ that yield higher $J_c$ in the layer and higher irreversibility field averaged over the whole A15 phase layer. Some conductors produced for ITER incorporated a large central source of tin or used bronze as the Sn source. The reduced tin activity of α-bronze works against achieving the higher layer $J_c$ properties of RRP and PIT, while the low solubility of Sn in α-Cu Sn bronze means that the fractional cross-section of A15 that can be formed with bronze conductors is about half that in RRP and PIT. Thus, high Sn conductors develop high $J_c$ both because of better layer properties and a higher volume fraction of the A15 in the overall reaction mixture.

Interesting exceptions have been summarized by Hopkins [28],[29] working in conjunction with CERN and manufacturers in Japan and Russia. Distributed-tin designs could advance after extended R&D to approach performance targets for high-field magnets. Continued support of research into alternative designs and in manufacturing capabilities worldwide will be required to alleviate supply chain challenges if production for future facilities is on the scale of that completed for ITER (which sourced from 8 manufacturers) and well beyond the scale of the Hi-Lumi LHC upgrade (which sourced from 1 manufacturer). Consolidation of manufacturing since the ITER procurement has reduced the overall manufacturing capacity and capabilities, however. At present no new architecture yet challenges RRP for overall conductor $J_c$, scale and supplied piece length. In our opinion, one of the biggest challenges of all high Sn routes is the complexity of the initial Sn-Cu mixing and then the various kinds of reaction layer that form while diffusing Sn form its source into the Nb.

*Advanced Nb alloys for Nb₃Sn conductors*

The metals in Groups IVa, Va (Nb is in group Va), and VIa form body-centered cubic alloys with significant mutual solubility. This condition permits adding elements that have strong affinity for oxygen (e.g. Zr, Hf) as well as elements with high atomic number (e.g. Ta, Hf, W). Development of Nb$_3$Sn tape over 30 years ago by General Electric [30], [31] utilized Nb1%Zr alloy coupled with anodization to create *in-situ* oxidation, by which ZrO$_2$ particles constrained Nb$_3$Sn grain growth during a subsequent high-temperature reaction. Virtually all Nb$_3$Sn conductors use grain boundaries as vortex pinning centers to drive up $J_c$. In addition to the much greater electromagnetic stability of bronze multifilament conductors, the ability to form the A15 phase at 650-700 C ensured that grain sizes of 100-200 nm were possible, not the micron-sizes of tapes formed above 930 C. Tapes disappeared from the market shortly after low temperature reacted, multifilament bronze conductors appeared [32] due to their excellent electromagnetic stability and much higher $J_c$. When accelerator builders took up the challenge of making Nb$_3$Sn suitable for accelerators, it was realized that bronze conductors could not generate sufficient critical current density and this stimulated much of the internal tin architectures that form the precursors for today's RRP & PIT conductors. Dietderich *et al.* [32] broadened the discussion by noting that oxide particles could not only constrain A15 grain growth but also form strong vortex pinning centers within Nb$_3$Sn grains. Under the stimulus of the challenges of an Energy Frontier Hadron Collider, a number of researchers [34], [35], [36], [37], [38] have subsequently explored the potential benefits of using internal oxidation to improve flux pinning and increase $J_c$ at high field in Nb$_3$Sn conductors. Conductor designs to achieve this goal will be discussed shortly.

The benefits of enhanced vortex pinning would be lost without concomitant retention or increase of the upper critical field. Here, addition of elements with high atomic number provides spin-orbit scattering [39], [40], [41] and suppression of any paramagnetic limiting, thereby enhancing the upper critical field. Tarantini *et al.* [42], [43] show the opportunities for increasing the upper critical field of Nb$_3$Sn conductors when alloyed with both Ta and Hf.

Because a central goal of all Nb$_3$Sn conductor development is to grow the A15 phase with grain size significantly smaller than 100 nm, these recent efforts focused attention on how the nucleation of the A15 phase occurs at the Nb-Sn reaction front. An important new observation in [37] is that addition of 1at.%Hf to Nb4at%.Ta raises the recrystallization temperature of the generally used pure Nb rods in standard RRP designs above that of the normal A15 reaction temperature. It appears that A15 nucleation occurs by diffusion of Sn into the alloy GBs, which allows A15 grains as small as 50 nm even in the absence of any oxide particles which confer similar benefits [34,35,36]. This new combination of grain refinement techniques is under active study at the present time.

Development of conductors by both RRP and tube routes is presently underway using advanced alloys in both university and national lab and industry. These developments demonstrate well the presence of a collaborative ecosystem linking initial discovery in universities with "nascent" wires made in university or small business to quasi-industry R&D. Greater conductor development resources could greatly accelerate the ability to ensure that real magnet conductors become available in the next 3-5 years. Special Nb alloy melting has been addressed by multiple raw material suppliers, and fabrication activity also spans multiple industry participants.

*Additions of high heat capacity materials into conductors*

Research from the Kurchatov Institute and Bochvar Institute in Moscow more than a decade ago demonstrated the ability of rare earth additives to add stability to magnet windings [44], [45] that could accelerate the rate of magnet training [46]. More recent considerations [47], [48] identify several challenges of Nb$_3$Sn magnets that could be alleviated by the addition of rare-earth oxides. As discussed earlier, the sub-element diameter of present RRP and PIT conductor types fall into a regime where magnetization instabilities can be a significant factor in initiating training in present Nb$_3$Sn dipoles. This challenge will become more difficult if $J_c$ is further increased using advanced conductor designs if $d_{\text{eff}}$ remains the same. Addition of a material with high heat capacity could offer an important stability counterbalance to the potential increase of sub-bundle magnetic energy by helping absorb and slow the energy released during flux jumps. Magnet stability and training can be addressed at the strand, cable, and coil level by further additions of materials with high heat capacity, e.g. via additives to epoxy or additions to the cable core. These are activities outside of the scope of this white paper.

Development of $Nb_3Sn$ conductors with high heat-capacity additives is underway in parallel with investigation of advanced alloys mentioned above. Challenges exist in identifying and then optimizing powders for both properties and compatibility with manufacturing.

> Statement: $Nb_3Sn$ conductors with advanced designs and improved performance at high field are presently in R&D. Pushing them into industrial pilot plant production so that magnet lengths of wire can be made within the next 3-5 years and will allow proper test of the magnet potential.

### Bi-2212 round wire conductor

Conductors made from high-temperature superconductor (HTS) are capable of much higher fields than any Nb-based conductor, as Figure 1 shows. Bi-2212 is quite unique as a cuprate HTS material in being able to develop high $J_c$ in a multifilament round wire form without macroscopic texture. Bi-2212 offers the potential to displace $Nb_3Sn$ above about 15 T field, see Figure 2, in round multifilament wire form. A key advantage of Bi-2212 over REBCO and Bi-2223 is that the processing that develops high $J_c$ passes through a state in which starting Bi-2212 powder melts to an almost complete liquid phase. Careful resolidification under conditions of sparse nucleation allows rapid growth of a few grains in the confined Ag filament tubes [49]. All cuprate superconductors require texture to permit current to pass from grain to grain; in the case of Bi-2212 made by powder-in-tube methods the melting step allows disconnected Bi-2212 phase powders that deform inside their Ag tubes by particle rolling to form connected dense liquid regions on melting. The problem of the melting though is that the residual 1/3 volume is principally air that suffers an ~4-fold increase in pressure that can swell or even burst the very soft Ag sheath. The oxygen can pass through the Ag but this is only 20% of the gas pressure since nitrogen remains trapped. During the slow solidification process that ensures sparse nucleation, large grains form randomly with low density. Fortunately, the 3 crystallographic axes have significantly different growth rates. Within filament tubes of ~15 μm diameter, the most rapid *a*-axis growth gobbles up off-axis grains and generates a strong [100] texture along the filament axis. The *b*-axis is orthogonal to *a* thus generating a biaxial texture of ~15°. A key issue of this aligned microstructure is that obvious weak link connections are suppressed, unlike Bi-2223, where only a uniaxial texture exists [51] [49], [50], [51].

The second important step in obtaining high $J_c$ is remedy the effects of entrained air in the powder and its tendency both to burst the wire during heat treatment (HT) and divide the filament into full density regions and voids or bubbles across which current flows only by bridging the voids by a few rapidly growing large Bi-2212 grains [51], [52]. The remedy is to perform the HT under sufficient pressure to prevent any effect of the internal gas pressure [53], [54] that would prevent full densification of the Bi-2212 filaments. Operation over the range 20-100 bar total pressure (1 bar is always oxygen with balance argon) allows >95% density of the Bi-2212. Because 2%$O_2$ in argon is standard welding gas, we have standardized to this formulation also for 50 bar overpressure HT.

Figure 5 summarizes the HT, the texture, and the final filament structure of an optimally processed Bi-2212 wire after such a 50 bar over-pressure HT. Such an OPHT for coils necessitates a wind-and-react magnet approach in specially designed apparatus at Florida State University shown in Figure 6.

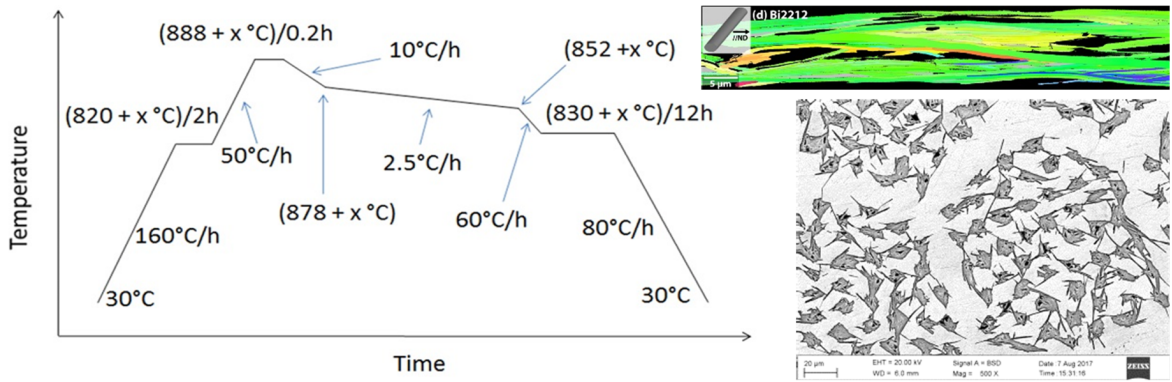

**Figure 5:** The melt-solidification process sequence and the resulting microstructure of Bi-2212 strand. A key aspect, partly related to the crystal structure of Bi-2212, is that crystal growth during slow re-solidification is much faster along the crystallographic *a* direction, which lies within the copper-oxide planes. When the filaments are well-defined and separated, as in the bottom right photo, this leads to the formation of macroscopic texture and alignment of the copper-oxide planes, as represented by the predominance of green color in the orientation-dependent image of a longitudinal cross section at top right. Texture is essential for avoiding obstruction to carrying current when anisotropic superconductors such as Bi-2212 are used.

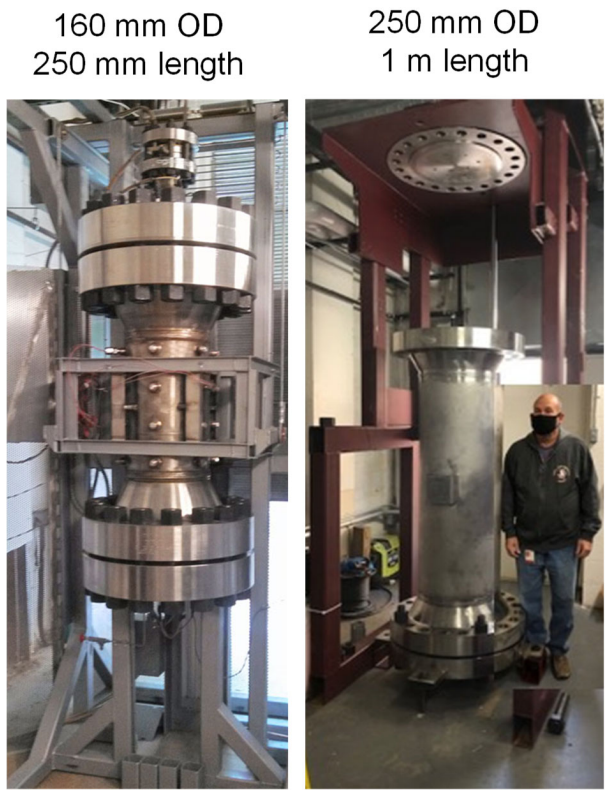

**Figure 6:** Reaction furnaces for Bi-2212 test coils operating at 50 bar pressure through the reaction sequences like that shown in Figure 5. Dimensions reflect the internal diameter and length of the uniform temperature zone. The larger furnace at right is being commissioned in 2022.

A round wire HTS architecture is important for the accelerator sector in several ways. First, wire-drawing techniques presently applied to copper-clad superconductors can be applied to Bi-2212 in large degree. This could permit scale-up and quality control like that essential to large-scale production of Nb-Ti and $Nb_3Sn$ conductors for accelerator magnets. Second, cabling techniques are generally unchanged between Nb-based conductors and Bi-2212. Third, although the need for twisting is not fully clarified, twisted conductors have been produced. This could provide an important advantage for ac loss in rotating machines, with possibly greater impact than the need to reduce losses upon ramping of accelerator magnets. Fourth, existing insulation winding and braiding schemes can be adapted to accommodate the temperature and chemical interactions for Bi-2212.

Many of the basic requirements for understanding Bi-2212 conductors and obstacles to their use in magnets were worked out by the Very High Field Superconducting Magnet Collaboration between 2009 and 2013, which combined HEP national laboratory teams with university groups. Further advances were achieved by advancing the production of Bi-2212 powder between 2013 and 2018 with a standard composition, the so-called "521" formulation [55]. Multiple, small-scale, manufacturers have produced good, fine powders that facilitate conductor fabrication with low breakage. Continued support from both DOE-HEP and the National Science Foundation (NSF) via the National High Magnetic Field Laboratory (NHMFL) has led to demonstrations of good reproducibility for coil reactions based on conductor witness samples.

New over-pressure reaction systems should allow dipole model coils up to 1 m length and 250 mm diameter to be completed by end of 2024, if provided adequate resources. This infrastructure should also facilitate exploration of laboratory solenoids and NMR systems at well above 30 T field. Infrastructure for longer magnets has received initial informal consideration.

### REBCO conductor

REBCO became a viable high-field magnet conductor when high-strength Hastelloy substrates became dominant [56], [57] and when advanced pinning center design showed how to develop extremely high layer current density in manufactured tapes [58], [59]. REBCO rapidly became the basis for high field magnets, especially the 32 T user magnet at the NHMFL [60], [61] and the Bruker Biospin 28.2 T, 1.2 GHz NMR system. Numerous other magnet demonstrations above 25 T have now been made as Table II summarizes. Such work, as for Bi-2212, has important bearing on the solenoids for a potential muon collider. Several projects are underway aiming toward 40 T magnets [62].

Further advances to reduce the Hastelloy thickness from a standard 50 μm to 30 μm [63] provide a significant enhancement of the winding current density, especially in the so-called "no insulation" condition [64]. This enhancement led to achievement of winding current densities of over 1400 A mm$^{-2}$ and reaching beyond 45 T field [65]. Stacked-tape cables [66], [67], [68] now make up a vibrant research field for fusion technology and potentially other applications for user and industry magnets. The scope of these manifold developments is too broad to summarize properly in this document.

Roebel cables were first developed for REBCO conductors about 15 years ago [69], [69], and they have been implemented in short model dipole coils [71]. Extensive characterization of Roebel cables [72] uncovered susceptibility to degradation under transverse stress, which could be partly managed by epoxy impregnation. A difficulty with the Roebel design is that wide REBCO tapes are required as starting material, and over half of the starting material is cut out and discarded by the patterning process, significantly driving up the cable cost. These challenges remain at the present time.

Thin Hastelloy is an enabler of compact round "wire" REBCO cables that provide an interesting alternative to Roebel cables. Conductor on round core (CORC®) [73] and symmetric tape round (STAR) [74] are two variants envisioned for future accelerator magnets. Each consists of a REBCO production tape conductor wound in a helical fashion around a conductive core. Thin substrates reduce the strain applied to the REBCO layer [75], which permits winding on smaller forms which enables smaller wire diameter and higher current density. Wires with typical diameter of ~ 3 mm are facilitated by the 30 μm Hastelloy product. Flexibility to bend either CORC® or STAR conductors is related to the tape width and winding pitch, where conductors that are narrower (now pushing below 2 mm width) and wound with a shorter helix pitch have the best flexibility. Unfortunately, these trends drive up cost by requiring more conductor per unit length of cored wire. At the present time losses due to damaged material at the slit edges is not negligible. Indeed development of laser slitting is proceeding fast at multiple companies, making gross damages implicated in magnet damage much less likely [63].

A key question for the future of REBCO coated conductor (CC) has perennially been: *When will the price come down to be attractive for electric utility users*? For more than a decade, the technical success of coated conductors in transmission cables, fault current limiters, motors and generators was thought to be a demonstration of replacement of many copper and iron machines with superconductors. The central dilemma was that coated conductors showed no possibility of getting to acceptable production levels and supposed low cost without commitments that were predictably unrewarding. All of this has potentially changed within the last two years as the case for tokamak fusion power plants operating with magnets with peak fields of 20 T at 20 K has gained huge private support. A prototype toroidal field coil used 500 km (about 1.5 ton) of 4 mm tape, well over 2 orders of magnitude more than the coated conductor put into *all* prior HTS magnets. The SPARC reactor being constructed will require 9,000 km (about 30 tons) more. Potentially therefore, there is now a very important feedback loop in place: Fusion magnet demands keep lowering the price of CC, allowing its replacement use for Cu and Fe for electro-technology. Such cost reductions have great promise for future colliders provided that suitable conductors for accelerator dipoles and quadrupoles can be developed.

Real project use has many benefits one of which is that routine testing generates important information about the stability of production. Such a program for the 11 km of tape purchased for the NHMFL 32 T user solenoid showed that many aspects of one company's production had significant property fluctuations of many types, in $J_c$, in dimensions, in Cu thickness and in damage generated by mechanical slitting. While some of these properties are tracked by the vendor, typically critical current properties are tracked only at 77 K and self-field. A central challenge is that the applications in the accelerator sector require verification of properties at 4 K, high field, and a range of angles, regimes in which it is much more difficult to carry out measurements due to the extremely high critical current and the associated risk for sample damage. Application of the so-called "lift factor" to estimate 4 K performance from 77 K breaks down when contributions from point pinning centers, e.g. oxygen vacancies, add to and compete with the correlated pinning due to artificial pinning centers, e.g. barium zirconate nanorods [76], [77], [78]. Significant variability in the 32 T conductor properties at 4 K have has been noted, even when the 77 K data shows reasonable uniformity [79]. Variations in properties have been connected to variations in the nanostructures, believed to be induced by fluctuations in growth conditions of the REBCO that are apparently beyond present process control both between different production runs and sometimes along the length of a single conductor. Some processing variations can lead to vulnerabilities in the magnet [80], [81]. In a sign that the highly optimized pinning centers possible in laboratory systems may not be replicable in 500–1000 m lengths, one important company has recently described a much simpler chemistry based only on Y, Ba, Cu and O where the dominant pinning centers are $Y_2O_3$ nanodisks generated by controlling the excess Y present during growth [82].

The discussion above points to a need for tight feedback between characterizations at 4 K and processing by the different manufacturers. While several discussions (some referenced above) have looked at the products coming off manufacturing lines, explicit optimization for 4 K and high field is only just beginning. A significant characterization effort is going into the manufacturing of cables, especially for fusion, and the research and development of very high field magnets. Several facilities around the world, notably at NHMFL, Robinson Research Institute, Tohoku University, and University of Geneva, are providing detailed characterizations of the strand for these cables in a field and angular range of interest for the accelerator sector. The vulnerability of cables to stress is receiving new attention in accelerator magnets [71],[83] and some fusion cable tests [84], [85], [86].

*Iron-based superconductors*

Iron based superconductors (FBS) present the opportunity of a material with very interesting primary properties and potential to become an important conductor. It is now in the delicate transition from the research lab to industry. Routes to make it as simple wires using the powder in tube route have been demonstrated [87], [88] but at present the best current densities are not compelling for applications, being about 100-300 A/mm$^2$ at 10 T 4.2 K.

FBS encompass several compound classes [89], [90], [91]. They are medium temperature superconductors with the interesting compounds having critical temperature $T_c$ lying in the 35-55 K range and extrapolated upper critical fields >50 T at 0 K (see Figure 1). In the rather isotropic so-called 122 compounds, of which $(K,Ba)Fe_2As_2$ may be the most interesting example, critical temperature $T_c$ is ~35 K and $H_{c2}(0)$ as high as 90 T with virtually no $H_{c2}$ anisotropy. The lack of anisotropy and the very high $H_{c2}$ are unequalled by any other superconducting compounds, setting up an immediately strong practical and scientific case for their use [92], [93]. Single crystals can have $J_c$ values exceeding $10^5$ A/mm$^2$ at 4.2 K and a few tesla. Sadly, early experiments [94], [95], [96] with [001] thin film bicrystals showed evidence for weak link behavior beyond about 5° misorientation, albeit with about a factor of 10 less angular $J_c$ sensitivity than in YBCO bicrystals. It is now accepted that there is an intrinsic weak link behavior in FBS.

However, early wires did demonstrate on the scale of magneto-optical imaging resolution, (~5 mm), that current flow was uniform [87] and that transport and magnetization (using the whole sample dimensions) measurements of $J_c$ agreed, all of which argues in favor of there being a true long-range supercurrent. The reasonable conclusion is that there are percolative supercurrent paths through the generally 100-500 nm grains that are much more effective than in comparable polycrystalline REBCO bulk or wire samples.

At this stage there is a strong Chinese program that has publicly announced a belief that FBS, especially $(K,Ba)Fe_2As_2$ can become a viable conductor and the group of Ma at CAS-Beijing has demonstrated 100 m long monofilament wires [97]. Their wires, actually tapes, do exhibit some mechanical texturing in rolling that appears to enhance $J_c$ in the favorable direction, but not yet to a level that is compelling for applications. In summary FBS materials are at the interesting stage where fine-grain polycrystalline compacts can be made as bulks or wires by PIT fabrication routes with current densities that approach the lower level of interest for applications (~100 A/mm$^2$ in 10 T at 4 K). The problem of not-yet-high-enough $J_c$ appears to be one of degraded percolative connectivity. Whether the connectivity is best enhanced by texture, by greater attention to purity (especially of grain boundaries) or managing of internal strain within the conductor (the $T_c$ of all FBS is very sensitive to the Fe-As bond angle) is not yet clear and is being actively studied by multiple groups [97], [98], [99], [100]. In short, the scientific case for researching how to enhance the connectivity of polycrystalline FBS remains strong because the route to wires is quite open if the appropriate materials processing route to higher connectivity can be understood. Moreover, their raw materials are inexpensive and routes to make wire conductors might be compatible with copper composite wire manufacturing.

## The role of public-private partnerships in conductor development

### Conductor Development Support under DOE-HEP

Superconducting wire development has been supported by DOE-HEP ever since the Tevatron, initially within the Fermilab, BNL and LBNL programs [101], [102] and then with the University of Wisconsin program [103]. Fermilab showed how to make Nb-Ti at scale for the Tevatron and exposed the fact that wire properties were variable and poorly understood. This attracted the attention of the Wisconsin group who saw the opportunity for improved wires flowing from a proper understanding of the process. The principal vehicle for stimulating the virtuous cycle of better understanding leading to higher $J_c$ leading to better wires was an annual workshop (The Nb-Ti Workshop first held in 1983, still running annually under the new title of the Low Temperature/High Field Superconductor Workshop). This workshop, now generally called LTSW, has always been small, typically 75 persons, to allow an unscripted discussion and interaction vehicle for the magnet users, the superconducting industry and the researchers trying to understand the pathways to better conductors. The Nb-Ti developments generated by this public-private partnership led not only to the SSC, RHIC and LHC conductors, but also to the huge tonnage industry of Nb-Ti for MRI magnets.

The success of this program then led to a new program whose goal was to develop new and much higher $J_c$ Nb$_3$Sn for accelerator magnets [104], [105], [106]. Under the Conductor Development Program (CDP), large manufacturers supplying Nb-Ti wire for the MRI industry, who were not eligible for incentives like those given to small businesses, were aided to innovate new wire products. Awards were based on proposals for innovative work and industry cost-share. This public-private partnership model led to key innovations of the RRP conductor in the years leading up to the Hi-Lumi LHC upgrade production run [1]. Combined investment of just under $4 million by LARP and CDP between 2005 and 2014 advanced the RRP configuration through 61, 127, and 169 stack variants as well as established production readiness [107]. While the scale of internal matching funds are not available to the public, a good estimate of the total *annual*

development investment in high-$J_c$ Nb$_3$Sn conductor is $1 million. This was followed by pre-production procurements of about 1 ton leading up to the Hi-Lumi LHC upgrade.

Unfortunately, the landscape for public-private partnerships has changed. Margins derived from medical imaging systems have decreased significantly since about 2000, which has practically eliminated the funds available for industrial cost share. The successor to CDP under DOE-HEP, the Conductor Procurement and R&D Program or CPRD, has broadened its scope to support Nb$_3$Sn, Bi-2212, and most recently REBCO, but has not been able to provide resources like those described above leading up to the Hi-Lumi LHC upgrade. As a result, advance of Nb$_3$Sn with new alloys and additives, as well as advance of Bi-2212 and REBCO as magnet conductors for the accelerator sector, has not proceeded as quickly as the program would like. Significant additional support does come under small business programs grants and research investigator awards, and these have been essential for supporting conductor development through the nascent wire stage.

### Nurturing the industry ecosystem

Real challenges lie in moving conductor development from nascent wires into industry R&D and pre-production. Present industry "pull" for high-performance Nb$_3$Sn wire is ~1 ton annually, which is orders of magnitude below production levels that would be needed for a major science facility. Importantly, the medical imaging market has not taken up Nb$_3$Sn conductors in a major way, and other markets for therapy and high-field NMR magnets do not generate as large a demand for advanced conductor as the MRI sector. The conclusion of the Nb$_3$Sn production run for the Hi-Lumi LHC upgrade could mark an important point in time, where investment decisions for Nb$_3$Sn, especially the premium RRP grades sought by accelerators, and Bi-2212 face a long, uncertain period until a possible future accelerator facility is launched, with a very modest marketplace in the interim. (We discuss REBCO in view of private funding for fusion below.) These considerations could force a shutdown of manufacturing in the US and consolidation of resources to other locations. Atrophy of expertise and capability would then follow, which would make the re-start period much longer and more expensive.

Continuity of the CPRD program is presently integrated within the goals of MDP. Actions to sustain and augment MDP and CPRD will be essential to continuing conductor development under the present mechanisms. This model is unlikely, however, to successfully compensate for the huge cost pressures felt by the wire industry from the MRI magnet manufacturers for whom Nb-Ti has become a commodity. It is important to note that the conductors desired by the accelerator sector generally require the highest performance: for example, Nb$_3$Sn for accelerator use has the highest $J_c$ of all routes. Although today's RRP wires do find application for many types of laboratory magnets, including the highest performance NMR magnet wires [108], manufacturing of strand for accelerators is always aiming for the highest properties that renders their goals somewhat decoupled from manufacturing to meet the needs of other customers. The recent end of the production run for the Hi-Lumi LHC upgrade places urgency on finding a better model for assuring conductor supply for a future science facility. This loss of demand, taken together with considerations such as manufacturing leases, anticipated conductor orders, and federal project profiles, could compel manufacturers to take actions that significantly impair US conductor manufacturing by 2024.

During 2022, supply-chain exercises in the DOE were initiated by an Executive Order [109], which propagated to various workshops and reports [110]. Funds were provided by the DOE Office of Accelerator R&D and Production (ARDAP) to explore the nature of public-private partnerships that could alleviate magnet conductor supply-chain risks and better assure availability of production for the accelerator sector. A workshop held 13-14 March 2022 at Tufts University debated many aspects of a potential business plan going forward. Among the elements thought to be essential for the plan are:
- <u>Sustained magnet development programs</u> like those driven by the accelerator sector, e.g. LARP and programs proposed for consideration by HEP [111]. Magnet development programs provide a virtuous cycle of defining ever more challenging targets and goals, pushing industry to develop ever better conductors in large quantities, making and testing lots of magnets (sometimes to the point of failure), and exposing and solving issues that underpin the advancing technology.

Programs with coupling to solenoids and other magnet configurations used by industry applications would especially be beneficial to the accelerator sector – industry ecosystem. R&D toward muon colliders are especially exciting for this reason.
- Sustained annual acquisitions of conductor over and above R&D wires to keep manufacturing "warm" at a pre-production level. Acquisitions of raw material, e.g. advanced Nb-Ta-Hf alloy, needed to allow emerging ideas to enter industrial R&D production could be very valuable and are presently being nurtured at a low level by CDRP. Pre-production levels for $Nb_3Sn$ would be 1 ton per year with configurations meeting standard specifications, e.g. the Hi-Lumi LHC upgrade specification. The configurations should anticipate future use by the accelerator sector, where for instance 1.0 mm diameter wires might be favored over 0.85 mm diameter.
- Managing a limited conductor stockpile as a national resource for both accelerator magnet R&D and growth of the commercial magnet industry. The stockpile would be limited to avoid inventories of obsolete conductors where R&D is rapidly moving. Development of new product lines and valley-of-death development challenges have been traditionally assisted by public funds, especially in technology areas of strategic interest. Assistance could involve withdrawals from the stockpile as prudent. Over the long term, development of market pull should emerge, creating a virtuous cycle like the development of Nb-Ti conductors.
- Nurturing workforce pipelines by supporting student internships in industry settings, industry traineeships at academic institutions, training partnerships, and post-graduate incentives. Besides universities, trade schools and junior colleges are seen as adding journeyman skills to the conductor and magnet manufacturing workforce.
- Coordinating an innovation institute to centralize a pool of manufacturing knowledge and establish a creative common around the key questions for conductor R&D and scale-up. Such an institute would build on the capabilities developed over 40 years by the culture of the Nb-Ti workshop and LTSW/HFSW. Entrepreneurial ideas that have strong merit often do not produce fruit for reasons related to limited resources or technical difficulties. An institute would coordinate access to capabilities and equipment, facilitate exploration of ideas through multiple attempts, encourage debugging of failures, and catalog methods with appropriate propriety. Legal teams involved in partnerships could be encouraged at early stages. Marketing and communication teams could also be engaged during early discussions.
- Incubate component development especially where components, such as superconducting solenoids, provide a strategic or critical advantage for larger systems. Products and components could be listed on the federal register.
- Incentivize industry ecosystem collaboration between national laboratories, universities, and industries. Activities to coordinate accelerator magnet manufacturing could be established in ways that cross boundaries of national laboratories and harvest the wealth of innovations in academia. Simplification of partnership mechanisms between national labs and industry would be helpful, such as via technical services frameworks and updates structures for collaborative research agreements.
- Provide industry access to measurements and test facilities at low cost. Quality plans associated with major procurements usually place the burden of quality control on the conductor supplier. However, significant time and expense is required to set up quality control capabilities, qualify them via benchmarks and inter-laboratory comparisons, and adequately carry out quality measurements. User facilities for industries at national laboratories could be helpful. For HTS conductors like REBCO, the present in-line capabilities often do not probe the conditions for end use of the product. Feedback between quality measurements and processing must be tightly looped, especially during R&D phases.

*Potential favorable disruptive factors in science and industry*

Close engagement of the HEP magnet development community with other areas of magnet research and development could prove to be very helpful for magnet conductor development.

Privately funded fusion energy research is acquiring REBCO conductor at the scale of a large science facility project, with ~$1 billion overall investment and a significant portion, ~$100 million, landing on conductor. This should drive several positive activities reminiscent of the virtuous cycle of development for Nb-Ti conductors:
- Magnet "pull" is being created by the vision and objectives of the fusion projects. A huge amount of credit is owed to the project leaders and their supporters.
- Multiple vendors are competing to provide conductor that meets a common specification. If the specification and conductor samples can be shared with publicly funded research groups, then a basis of important processing – structure – property links can be determined for each conductor, and research to reduce or eliminate property variations can begin. Coupling to lab and university groups could provide diagnosis and feedback to the manufacturers. Programs like INFUSE are already allowing a start on such collaborations.
- Techniques are being developed to measure $J_c$ properties at the field, field angle, and temperature of anticipated end use, besides the usual manufacturer characterizations at 77 K and self-field. The related measurement science, e.g. that performed for high-field magnet projects such as the 32T user magnet at NHMFL [79], is undergoing continuous activity in response. The high current density of REBCO makes transport measurements especially vulnerable to burn-out, and magnetometry is emerging as an alternative technique [112]. Better networking between the laboratories, universities, and private projects could improve the capability of manufacturers and partner labs to provide quality assurance relevant for end-use conditions.
- Many small test magnets are being manufactured and tested to conductor limits, but there is still much confidentiality associated both with successes and failures of HTS magnets. An important lesson of the Nb-Ti Workshop is that manufacturers originally came to the workshop with a view that they each had an optimized process that was proprietary and special. The strong interactivity of presentation and discussion at the workshops soon lead to entirely new ways of looking at optimization and the ultimate potential of Nb-Ti, leading to the recognition that sharing of experience drove an ultimately much more rewarding technology. If the results of present magnet testing within the various laboratories and research groups could also be shared, broad discussion could occur about potential conductor vulnerabilities exposed by tests. These discussions pull the materials characterizations and help define the topics for PhD research in university groups.

Superconducting rotating machines for sustainable energy are presently entering the marketplace. Many activities worldwide aim to develop wind turbines in the 15–25 MW range [113], [114]. Propulsion for ships, trains, and airplanes is also under development based on the significant increase in power density supplied by high-field superconducting coils. For example, electric aircraft are envisioned to significantly exceed 10 kW / kg and displace high-bypass turbofans as the propulsion unit [115]. Racetrack field coil configurations in many machines have synergy with dipole magnets in accelerators, so there could be opportunity for public-private partnerships. Many envisioned applications operate at temperature higher than 10 K, for which HTS and possibly iron-based superconductors are attractive. A 10 GW wind farm could require procurement of 100 tons of conductor.

Changes in medical diagnosis and therapy guidelines could drive increased development of high-field imaging systems, particle-beam therapy systems, and high-field NMR tools for *in vivo* characterization. Industry has begun to put 30-tesla class NMR systems into service, which require HTS conductors, but results from using these systems is only beginning to disseminate to the medical community. Industry has also begun to deploy compact synchrotron sources using high-field magnets on gantries, but these are in early stages of market penetration and the business drivers are still clarifying. World-class facilities like ISEULT have just begun courses of scientific research relevant for diseases of the brain and other potential therapies where extremely high resolution is needed.